\documentclass[acmsmall,screen,nonacm]{acmart} 

\usepackage{enumitem}
\usepackage{tabularx}
\setlist{nosep, leftmargin=*}
\usepackage{xcolor}
\usepackage{graphicx}
\usepackage{subcaption}
\usepackage{booktabs}
\usepackage[most]{tcolorbox}
\usepackage{makecell}
\usepackage{adjustbox}

\setcopyright{acmlicensed}
\copyrightyear{2026}
\acmYear{2026}
\acmDOI{XXXXXXX.XXXXXXX}

\begin{document}

\title{Supporting Calibrated Reliance in AI-Assisted Decision-Making: Different Strategies for Different Tasks}


\author{Ruth Cohen}
\affiliation{%
  \institution{Bar-Ilan University}
  \city{Ramat-Gan}
  \country{Israel}
}
\email{ruticohen770@gmail.com}

\author{Lu Feng}
\affiliation{%
  \institution{University of Virginia}
  \city{Charlottesville}
  \country{USA}}
\email{lu.feng@virginia.edu}

\author{Ayala Bloch}
\affiliation{%
  \institution{Department of Psychology, Ariel University}
  \city{Ariel}
  \country{Israel}}
\email{ayalablo@ariel.ac.il}

\author{Sarit Kraus}
\affiliation{%
  \institution{Bar-Ilan University}
  \city{Ramat-Gan}
  \country{Israel}
}
\email{sarit@cs.biu.ac.il}



\begin{abstract}
As AI systems increasingly support human decision-making, a central challenge is determining what information helps people recognize when to rely on AI predictions and when to question or override them. Across three controlled human-subject studies spanning abstract visual reasoning with RAVEN matrices and deductive logical reasoning with LSAT problems, we examine how different forms of AI support affect performance in AI-assisted decision-making. A multi-stage reveal study shows that AI predictions and explanations can affect objective accuracy and subjective confidence differently. In visual reasoning, LLM explanations do not improve accuracy beyond the predicted answer alone, and no additional support format significantly outperforms prediction-only support; predicted probabilities show the highest descriptive accuracy and error recovery, while a derived selective-automation policy provides a higher-performing reference benchmark. In language-based logical reasoning, by contrast, LLM explanations yield the highest accuracy and error recovery, outperforming expert-written explanations and probability-based support. These results suggest that the effectiveness of AI support strategies depends on the task setting. Human--AI interfaces should instead be designed to support calibrated reliance and effective error recovery by matching the form of assistance to the task and the evidence available to users.
\end{abstract}




\newcommand{\sectsectref}[2]{Sections~\ref{#1} and \ref{#2}}
\newcommand{\sectref}[1]{Section~\ref{#1}}
\newcommand{\figref}[1]{Figure~\ref{#1}}
\newcommand{\tabref}[1]{Table~\ref{#1}}
\newcommand{\red}[1]{\textcolor{red}{#1}}
\newcommand{\blue}[1]{\textcolor{blue}{#1}}

\maketitle

\section{Introduction}

As AI systems become increasingly integrated into human decision-making workflows, a central challenge is not simply whether AI can make accurate predictions, but whether people can determine \emph{when to rely on those predictions and when to question or override them}. Effective AI-assisted decision-making therefore requires interaction strategies that help users evaluate AI outputs and calibrate their reliance.

Explainable AI (XAI) has long sought to support such evaluation through methods ranging from post-hoc interpretability techniques such as LIME~\cite{ribeiro2016why} and SHAP~\cite{lundberg2017unified} to visual methods such as Occlusion Sensitivity (OS)~\cite{zeiler2014visualizing}. More recently, Large Language Models (LLMs) have enabled fluent natural-language explanations accessible to non-experts~\cite{naveed2024overview,miller2019explanation}. Yet explanations do not necessarily improve human--AI team performance. Human--AI teams often fail to outperform their strongest individual component~\cite{bansal2021does,steyvers2022bayesian}, and explanations can provide little additional benefit beyond the AI prediction or increase over-reliance and automation bias~\cite{parasuraman1997humans,cummings2017automation}. Fluent but incorrect LLM rationales may further mask AI errors~\cite{spitzer2025don,stringham2025teaching}.

A related concern is that XAI is often evaluated using subjective measures such as trust, confidence, and perceived understanding~\cite{hoffman2018metrics}, which do not always predict objective decision quality~\cite{buccinca2020proxy}. 
Alternative forms of support, including predicted probabilities over the answer options, may sometimes better support calibrated reliance~\cite{rechkemmer2022confidence,bansal2021does}.
{Here, we define an AI support strategy as the type of information presented to users to assist their decision-making, ranging from the AI prediction alone to additional information such as natural-language explanations, visual explanations highlighting influential input features, or predicted probabilities conveying uncertainty across answer options.}
Different support strategies provide different forms of evidence and impose different demands on users, so their effectiveness may depend on the task and decision context.

In this article, we ask:
\emph{What forms of AI support help humans calibrate their reliance on AI, and how does their effectiveness vary across task settings?}
We address this question through three controlled human-subject studies spanning abstract visual reasoning with RAVEN matrices~\cite{zhang2019raven} and deductive logical reasoning with LSAT questions~\cite{yu2020reclor}.
Across these studies,\footnote{The studies were approved by the Human Subjects Institutional Review Board at Bar-Ilan University, approval no.~071124338. Experimental materials, anonymized participant-level data, and analysis resources are available at \url{https://github.com/RuthiCohen/calibrated-reliance-human-ai}.}
we evaluate how different forms of AI support affect accuracy, reliance, error recovery, and subjective perceptions.

Our results show strong task dependence. In visual reasoning, the AI prediction provides the primary accuracy gain, with no significant differences among the additional support formats; predicted probabilities show the highest descriptive accuracy and error recovery, while a derived selective-automation policy provides a higher-performing reference benchmark. In language-based logical reasoning, by contrast, LLM-generated explanations provide the strongest performance. These findings suggest that effective {AI-assisted decision-making} requires matching the form of support to the task rather than assuming that any single strategy is universally effective.

\subsection{Key Insights}

\begin{itemize}
    \item Effective {AI-assisted decision-making} depends not only on model accuracy, but also on whether users can determine when to rely on AI outputs and when to override them.
    \item The effectiveness of AI support is task-dependent, suggesting that different decision contexts may require different support strategies.
    \item Evaluating {AI-assisted decision-making} requires attention to objective performance and error recovery, not only subjective measures such as confidence and trust.
\end{itemize}


\section{The Anatomy of a Human-AI Decision}
\label{sec:study1}

To understand how AI support shapes human reliance, we first disentangle the
effect of an AI system's prediction from the additional effect of its explanation.
Explanations are often intended to help users assess AI recommendations, but it
remains unclear whether they improve decision quality beyond the prediction
itself or instead primarily affect users' confidence. Our first study therefore
uses a multi-stage reveal design that isolates the effects of predictions and
explanations within the same decision process.

\begin{figure}
\centering
\includegraphics[width=0.8\linewidth]{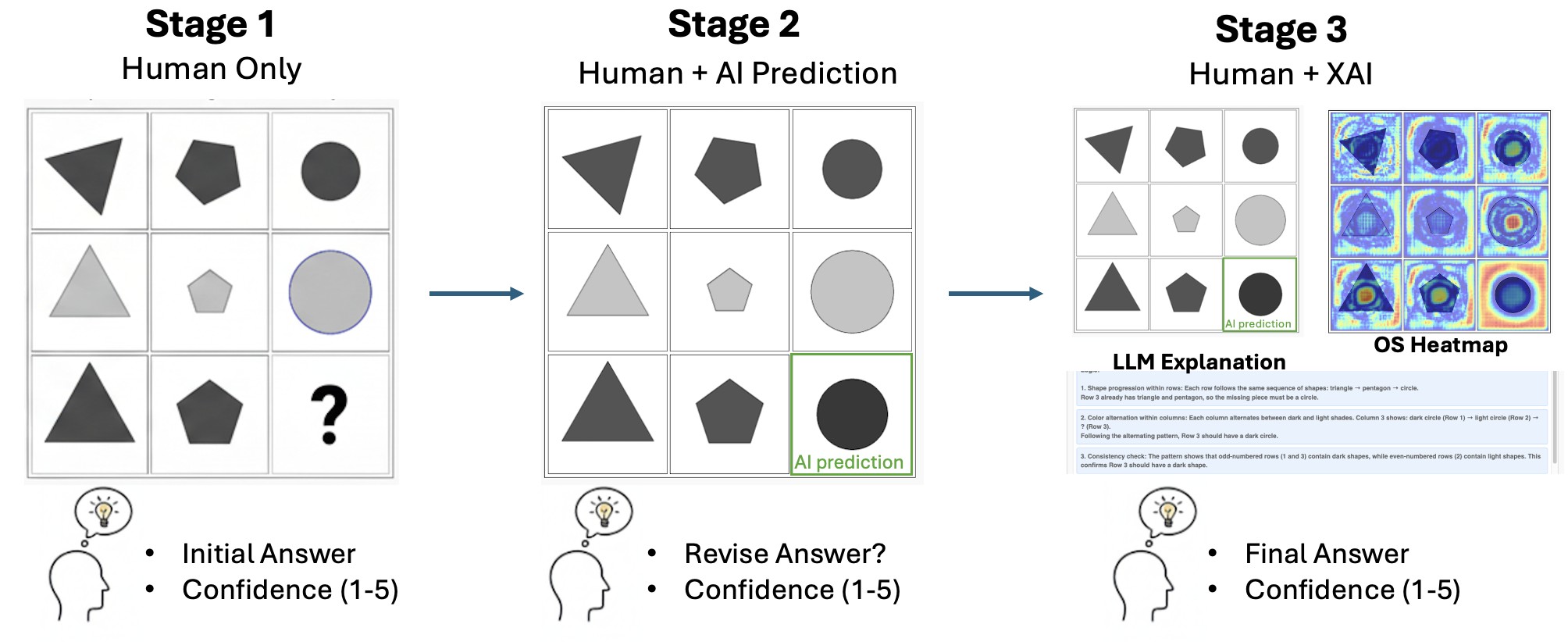}
\caption{The multi-stage human--AI decision process used to separate the effects
of the AI prediction and its explanation on user decisions.}
\label{fig:multi-stage}
\end{figure}

\subsection{Study Design: Within-Subjects Multi-Stage Reveal}

We examine abstract visual reasoning using RAVEN matrices~\cite{zhang2019raven}, which require inferring a missing panel through relational reasoning. 
These problems require multi-step logic, making them well suited for examining how different
forms of AI assistance influence human decision-making.

Participants ($n = 27$) solved up to 16 RAVEN puzzles in two blocks. 
Participants received performance-based compensation tied to the
number of correctly answered puzzles, providing a direct incentive for accuracy.
Recruitment, demographics, consent details are reported in Appendix~\ref{app:participants}.

In each block, the AI's predicted answer was accompanied by one of two
explanation types: a visual occlusion-sensitivity (OS) heatmap or a
natural-language rationale generated by an LLM. Within each block, the prediction
and explanation came from the same model: Claude 3.7 Sonnet in the LLM block and
DCNet~\cite{zhuoeffective} in the OS block. Additional model and explanation-generation details are
provided in Appendix~\ref{app:study1}.

To isolate the contribution of each layer of AI support, participants completed
each puzzle using a multi-stage reveal procedure:
\begin{itemize}
    \item \textbf{Stage 1 (Before AI):} Participants selected an answer and
    reported confidence on a 1--5 Likert scale without AI assistance.
    \item \textbf{Stage 2 (After Prediction):} The AI's predicted answer was
    revealed, allowing participants to revise their answer and confidence.
    \item \textbf{Stage 3 (After Explanation):} An LLM-generated explanation or
    an OS heatmap was presented, followed by a final opportunity to update the
    answer and confidence.
\end{itemize}

We measured objective accuracy and self-reported confidence at each stage.
Figure~\ref{fig:multi-stage} illustrates the three-stage reveal procedure,
which allows us to distinguish changes attributable to the AI prediction from
additional changes following the explanation.

Participants were informed that the AI system had an approximate historical
accuracy of 76\%, as displayed in the experimental interface. This information
was intended to convey that the AI was capable but fallible; the derivation of
this value is reported in Appendix~\ref{app:reported-accuracy}. 

\begin{figure}
\centering
\includegraphics[width=0.5\linewidth]{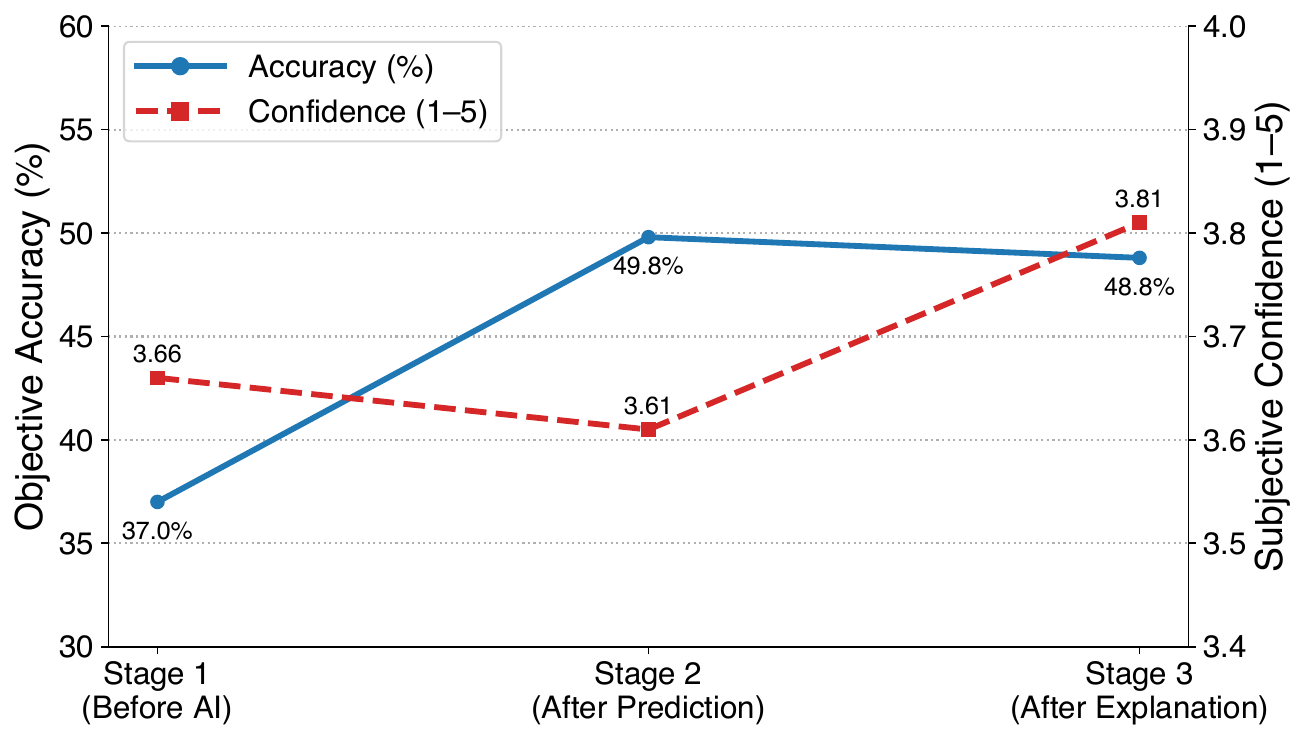}
\caption{Overall accuracy and self-reported confidence across the three reveal
stages in the multi-stage RAVEN study. Accuracy improves after the AI prediction
but plateaus once explanations are introduced, while confidence increases only
after explanations are presented.}
\label{fig:study1}
\end{figure}

\subsection{Results: Accuracy--Confidence Misalignment}

We first report overall accuracy and confidence aggregated across explanation
conditions to characterize the effects of introducing AI predictions and
explanations.

Figure~\ref{fig:study1} shows a clear divergence between objective accuracy and
subjective confidence across the three reveal stages. Participants' baseline
accuracy in Stage~1 was low (37.0\%), reflecting the difficulty of the task
without assistance. Accuracy increased substantially after the AI's predicted
answer was revealed in Stage~2 (49.8\%), although it remained well below the
AI's approximate historical accuracy of 76\% reported to participants. 
Explanations did not yield additional gains in Stage~3 (48.8\%). 
In contrast, confidence remained largely
unchanged after the prediction stage and increased only after explanations were
presented (mean confidence: 3.66 in Stage~1, 3.61 in Stage~2, and 3.81 in
Stage~3). Despite this increase in confidence, explanations did not improve task
accuracy beyond what was achieved by the AI prediction alone, indicating a
mismatch between participants' subjective confidence and objective decision
quality.

These effects were supported by Friedman tests across stages for both accuracy
and confidence (accuracy: $\chi^2(2) = 37.98$, $p < .001$, Kendall's $W = 0.70$;
confidence: $\chi^2(2) = 7.10$, $p = .029$, $W = 0.13$). Post hoc Wilcoxon
signed-rank tests with Holm correction showed a significant improvement in
accuracy from Stage~1 to Stage~2 ($p < .001$), but no difference between the
prediction and explanation stages ($p = .960$). In contrast, confidence
increased significantly from Stage~2 to Stage~3 following the introduction of
explanations ($p = .010$).

We also examined whether the incremental effect of explanation exposure depended
on whether the preceding AI prediction was correct. We found no significant
moderation by AI-prediction correctness; full analyses are reported in
Appendix~\ref{app:moderation}.

{We conducted an exploratory analysis using logistic regression with participant-level clustering to examine how initial confidence relates to AI reliance. Higher pre-AI confidence was associated with a lower likelihood of adopting a conflicting AI recommendation (odds ratio [OR] = 0.58, 95\% confidence interval [CI]: 0.45--0.75, $p < .001$), corresponding to approximately 42\% lower odds for each one-point increase in confidence. This relationship did not significantly differ between the two explanation conditions (interaction OR = 0.94, $p = .821$). See Appendix~\ref{app:confidence} for details.}


\section{Comparing Forms of AI Support} \label{sec:study2}

Motivated by the pattern observed in the multi-stage reveal study, we conducted a single-stage, between-subjects study to evaluate which forms of AI support most effectively improve objective task performance. This study moves beyond isolating the effects of explanations to directly compare alternative support strategies under identical task conditions.

\subsection{Study Design: Between-Subjects Comparison}

Participants ($n=100$) were randomly assigned to one of five human--AI support conditions, with 20 participants assigned to each condition:
\begin{itemize}
    \item \textbf{Human Only:} Participants solved puzzles without any AI support.
    \item \textbf{Prediction Only:} Participants viewed the AI's predicted answer.
    \item \textbf{Prediction + LLM Explanation:} Participants viewed the AI's prediction accompanied by a natural-language explanation generated by an LLM.
    \item \textbf{Prediction + OS Heatmap:} Participants viewed the AI's prediction accompanied by a visual occlusion-sensitivity (OS) heatmap.
    \item \textbf{Prediction + Probability:} Participants viewed the AI's prediction together with its predicted probability distribution across all answer options.
\end{itemize}

The AI source varied by support condition. Claude 3.7 Sonnet provided both predictions and explanations in the LLM-explanation condition, while DCNet provided the predictions and accompanying information in the OS-heatmap and predicted-probability conditions. The Prediction Only condition included an equal mix of predictions from Claude 3.7 Sonnet and DCNet. Additional methodological details are provided in Appendix~\ref{app:study2}.

Each participant solved 10 challenging RAVEN puzzles. To hold AI accuracy constant across the AI-support conditions, we selected items such that each condition contained six correct and four incorrect AI predictions, corresponding to an AI accuracy of 60\%. The model outputs themselves were not altered; the target accuracy was achieved through item selection.

For each puzzle, participants selected an answer, with objective accuracy recorded at the trial level. Participants in conditions providing additional information beyond the AI prediction also reported perceived understanding, clarity, and trust using five-point Likert scales.

We additionally analyze two decision-level measures that characterize participants' responses to correct and incorrect AI predictions. \emph{Agreement with correct AI predictions} is defined as the proportion of trials in which a participant selected the AI's answer when the AI prediction was correct. \emph{Error recovery rate} is defined as the proportion of trials in which a participant rejected the AI's answer and selected the correct solution when the AI prediction was incorrect.

We also report a post-hoc selective-automation benchmark that accepted the AI prediction when the gap between its two highest predicted probabilities exceeded a threshold and otherwise retained the participant's answer. The policy was evaluated using the predicted-probability condition and was not an independent participant group; details are provided in Appendix~\ref{app:selective-policy}.

\begin{figure}
    \centering
    \includegraphics[width=0.4\linewidth]{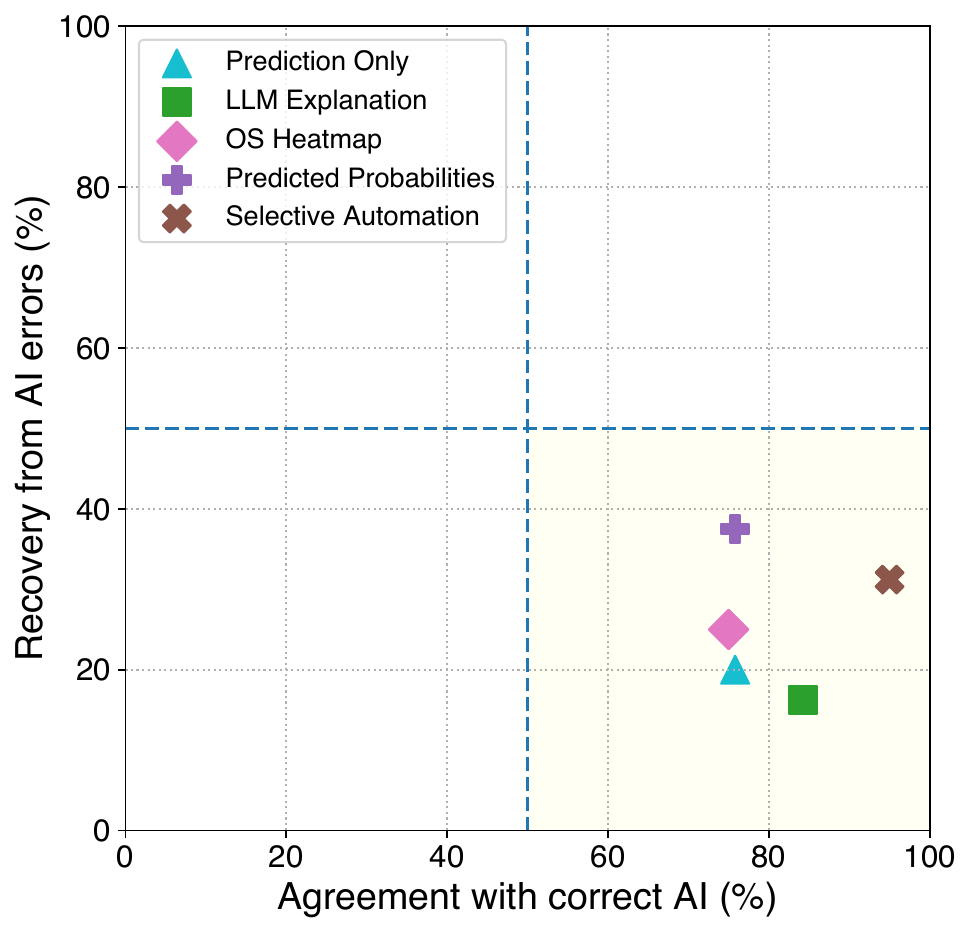}
    \caption{Relationship between agreement with correct AI predictions and recovery from incorrect AI predictions across AI-support conditions in the RAVEN task. The selective-automation policy is a derived benchmark based on responses from the predicted-probability condition rather than an independent participant group.}
    \label{fig:study2_recovery}
\end{figure}

\begin{figure}
    \centering
    \includegraphics[width=0.5\linewidth]{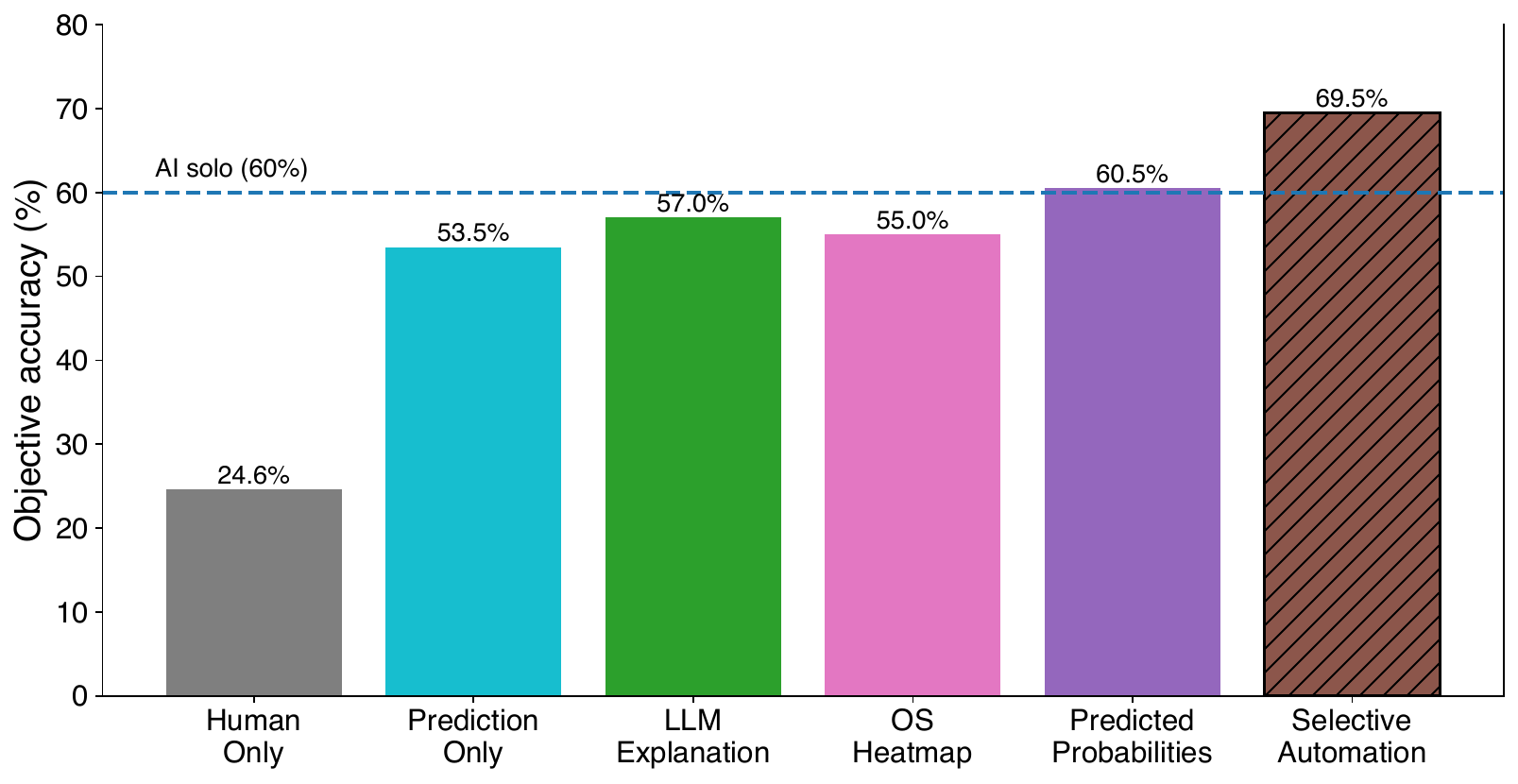}
    \caption{Objective accuracy across the five participant-facing conditions in the RAVEN task, with the derived selective-automation policy shown as a reference benchmark.}
    \label{fig:study2_accuracy}
\end{figure}

\subsection{Results: Reliance and Performance Across Support Conditions}

\paragraph{Reliance and Error Recovery.}

\figref{fig:study2_recovery} plots agreement with correct AI predictions against recovery from incorrect predictions. Across all AI-support conditions, agreement was substantially higher than recovery. Descriptively, LLM explanations showed the lowest recovery rate (16.2\%) despite high agreement (84.2\%), whereas the predicted-probability condition showed the highest recovery (37.5\%) with 75.8\% agreement. However, differences across conditions were not statistically significant for either recovery (Kruskal--Wallis $H(3)=7.59$, $p=.055$, $\varepsilon^2=0.060$) or agreement ($H(3)=4.53$, $p=.210$, $\varepsilon^2=0.020$).

The derived selective-automation benchmark achieved 95.0\% agreement and 31.2\% recovery. Because it reuses responses from the predicted-probability condition, it was excluded from the between-subjects statistical comparisons.

\paragraph{AI Support Improves Task Accuracy.}

\figref{fig:study2_accuracy} shows that accuracy increased from 24.6\% for Human Only to 53.5--60.5\% across the four AI-support conditions. Overall differences were significant (Kruskal--Wallis $H(4)=33.95$, $p<.001$, $\varepsilon^2=0.315$). Post hoc Dunn tests with Holm correction showed that Human Only performed significantly worse than each AI-support condition (all $p<.001$, $|\text{Cliff's }\delta|=0.74$--$0.87$), while none of the four AI-support conditions differed significantly from one another (all adjusted $p=1.000$).

Thus, the AI prediction provided the primary performance gain, with no statistical evidence that explanations, heatmaps, or predicted probabilities further improved accuracy. The predicted-probability condition was descriptively highest (60.5\%), while the derived selective-automation benchmark achieved 69.5\%; the latter is reported descriptively because it is not an independent participant condition.

\paragraph{Probability Interfaces Are Rated Most Favorably.}

\tabref{tab:subjective} (left) reports subjective ratings for conditions providing information beyond the AI prediction. Ratings differed significantly for clarity (Kruskal--Wallis $H(2)=7.53$, $p=.023$, $\varepsilon^2=0.097$), but not for understanding or trust. The predicted-probability condition received the highest mean ratings on all three measures, with significantly higher clarity ratings than the explanation-based conditions ($p=.034$, Cliff's $\delta=0.45$).

\begin{table}
  \centering
  \begin{tabular}{lccc|ccc}
    \toprule
    & \multicolumn{3}{c}{RAVEN Study} & \multicolumn{3}{c}{LSAT Study} \\
    \cmidrule(lr){2-4} \cmidrule(lr){5-7}
    & LLM & OS & Probability & LLM & Expert & Probability \\
    \midrule
    Understanding & 4.10 & 3.52 & \textbf{4.21} & \textbf{4.40} & 3.85 & 3.26 \\
    Clarity       & 3.50 & 3.43 & \textbf{4.42} & \textbf{4.40} & 3.70 & 3.74 \\
    Trust         & 3.95 & 3.67 & \textbf{4.54} & \textbf{4.45} & 3.80 & 4.00 \\
    \bottomrule
  \end{tabular}
\caption{Mean subjective ratings (1--5 Likert scale) of perceived understanding, clarity, and trust in the RAVEN and LSAT studies. Ratings are reported only for conditions that provided additional information beyond the AI prediction.}
  \label{tab:subjective}
\end{table}


\section{The LSAT Study} \label{sec:study3}

To examine whether the effectiveness of AI support generalizes beyond visual reasoning, we conducted a study using logical reasoning questions drawn from the LSAT~\cite{yu2020reclor}. Unlike RAVEN matrices, these LSAT problems rely on verbal logic and structured argumentation, providing a complementary domain in which narrative explanations may play a different role in {AI-assisted decision-making}.

\subsection{Study Design: Between-Subjects Comparison}

Participants ($n = 80$) were recruited and randomly assigned to one of four between-subjects conditions, with 20 participants assigned to each condition:

\begin{itemize}
    \item \textbf{Prediction Only:} Participants viewed only the AI's predicted answer.
    \item \textbf{Prediction + LLM Explanation:} Participants viewed the AI's prediction accompanied by a natural-language rationale generated by an LLM.
    \item \textbf{Prediction + Expert Explanation:} Participants viewed the AI's prediction accompanied by an expert-written rationale.
    \item \textbf{Prediction + Probability:} Participants viewed the AI's prediction along with its predicted probability for each answer option.
\end{itemize}

The AI predictions, LLM explanations, and predicted probabilities were generated using ChatGPT-5, while the expert explanations were adopted from Bansal et al.~\cite{bansal2021does}. Generation and source details, as well as representative interfaces for the LLM-explanation and predicted-probability conditions, are provided in Appendix~\ref{app:study3}.

Each participant solved ten LSAT questions. As in the RAVEN study described in Section~\ref{sec:study2}, model accuracy was held constant at 60\% across conditions. Participants selected an answer for each question, with objective accuracy and subjective ratings recorded for each trial.

\begin{figure}
    \centering
    \includegraphics[width=0.35\textwidth]{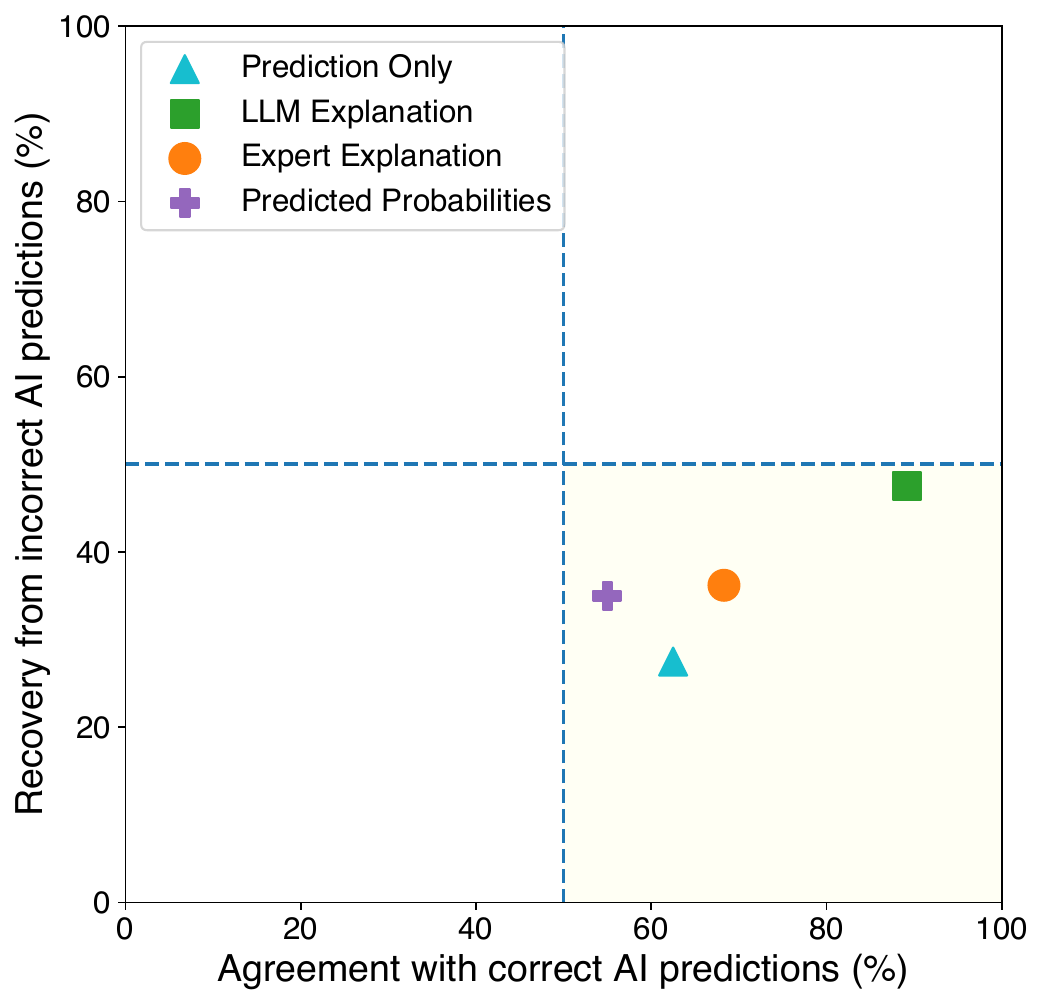}
    \caption{Relationship between agreement with correct AI predictions and recovery from incorrect AI predictions across support conditions in the LSAT task.}
    \label{fig:lsat_agreement_recovery}
\end{figure}

\begin{figure}
    \centering
    \includegraphics[width=0.4\textwidth]{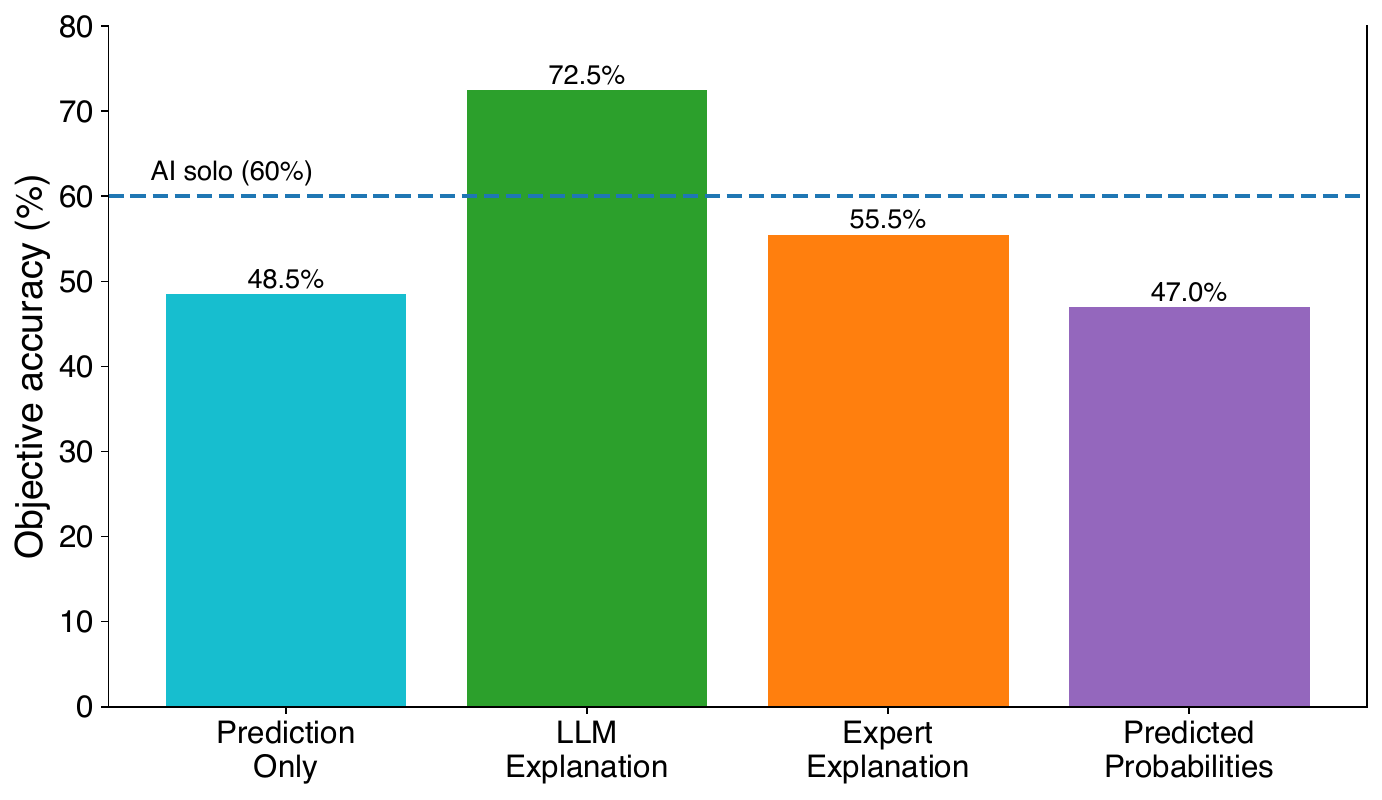}
    \caption{Objective accuracy across human--AI support conditions in the LSAT task.}
    \label{fig:lsat_accuracy}
\end{figure}

\subsection{Results: LLM Explanations Perform Best on the LSAT Task}

\paragraph{LLM Explanations Increase Both Reliance and Error Recovery.}
Figure~\ref{fig:lsat_agreement_recovery} plots agreement with correct AI predictions against recovery from incorrect AI predictions across support conditions in the LSAT task. Overall recovery rates differed significantly across conditions (Kruskal--Wallis $H(3) = 22.30$, $p < .001$, $\varepsilon^2 = 0.254$).

The LLM explanation condition exhibited both the highest agreement with correct AI predictions (89.2\%) and the highest recovery rate from incorrect AI predictions (47.5\%), indicating that LLM-generated rationales supported both effective use of correct model outputs and improved detection of AI errors. Expert explanations showed lower agreement (68.3\%) and recovery (36.2\%) than LLM explanations, compared with prediction-only support (agreement: 62.5\%, recovery: 27.5\%) and predicted probabilities (agreement: 55.0\%, recovery: 35.0\%).

Despite achieving the highest recovery rate, recovery in the LLM explanation condition remained below 50\%. Thus, participants failed to correct more than half of the AI's incorrect predictions even while agreeing with 89.2\% of its correct predictions, indicating that substantial error masking remained even in the best-performing support condition.

\paragraph{LLM Explanations Yield the Highest Task Accuracy.}
Figure~\ref{fig:lsat_accuracy} compares objective task accuracy across support conditions, with overall differences confirmed by a Kruskal--Wallis test ($H(3) = 22.30$, $p < .001$, $\varepsilon^2 = 0.254$).

The LLM explanation condition achieved the highest accuracy (72.5\%), substantially outperforming prediction-only support (48.5\%), expert explanations (55.5\%), and predicted probabilities (47.0\%). Post hoc Dunn tests with Holm correction confirmed significant advantages over all other participant-facing conditions ($p < .001$, Cliff's $\delta = 0.70$--$0.74$). Notably, the LLM explanation condition was the only participant-facing condition to exceed the model's standalone accuracy of 60\%.

\paragraph{LLM Explanations Are Rated Most Favorably.}
Table~\ref{tab:subjective} reports participants' subjective ratings of perceived understanding, clarity, and trust on the LSAT task. Ratings differed significantly across conditions for perceived understanding (Kruskal--Wallis $H(3) = 11.32$, $p = .010$, $\varepsilon^2 = 0.110$) and clarity ($H(3) = 10.76$, $p = .013$, $\varepsilon^2 = 0.102$), but not for trust.

The LLM explanation condition received the highest ratings for perceived understanding, clarity, and trust (4.40, 4.40, and 4.45, respectively). Post hoc Dunn tests with Holm correction confirmed higher clarity ratings than both expert explanations and prediction-only support ($p = .012$, Cliff's $\delta = 0.50$). In this study, these subjective ratings were consistent with the objective performance results: the LLM explanation condition also achieved the highest accuracy and strongest error recovery among the four support conditions.
\section{Discussion}

\subsection{Calibrated Reliance and Error Recovery}

Across the three studies, the effectiveness of AI support depended not only on whether participants followed correct AI predictions, but also on whether they could recognize and override incorrect ones. In the multi-stage RAVEN study, revealing the AI prediction improved accuracy, whereas adding an explanation increased confidence without a further accuracy gain. The between-subjects studies further showed that agreement with correct predictions and recovery from AI errors varied across tasks and support formats.

In RAVEN, LLM explanations showed high agreement with correct predictions but the lowest descriptive recovery from incorrect ones, whereas predicted probabilities showed the highest descriptive recovery; however, these differences were not statistically significant. In LSAT, LLM explanations supported both the highest agreement and the highest error recovery. {AI-assisted decision-making} should therefore be evaluated not only through overall accuracy or subjective perceptions, but also through responses to correct and incorrect AI outputs separately.

{Our exploratory analysis suggests that users' initial confidence may influence their willingness to follow conflicting AI recommendations. However, this association does not necessarily indicate more calibrated reliance, as resisting AI advice can be beneficial when the AI is incorrect but detrimental when it is correct. Future work could systematically examine how initial confidence interacts with different AI support strategies to influence appropriate reliance and error recovery.}

\subsection{Task-Dependent Effectiveness of AI Support}

The contrast between RAVEN and LSAT suggests that the effectiveness of additional AI support depends on the task setting. In RAVEN, providing an AI prediction substantially improved accuracy over unaided human performance, but adding explanations, heatmaps, or predicted probabilities did not yield significant further gains. Predicted probabilities nevertheless showed the highest descriptive accuracy and error recovery. In LSAT, by contrast, LLM-generated explanations achieved the strongest objective accuracy, error recovery, and subjective ratings.

Several non-mutually-exclusive factors may contribute to this difference. One is representational alignment: RAVEN relies heavily on visuospatial processing and relational integration~\cite{carpenter1990one,jung2007parieto}, so translating verbal explanations into spatial representations may impose additional cognitive demands~\cite{sweller1988cognitive,baddeley2000episodic}. LSAT questions and explanations, by contrast, share a language-based representational format that can make argument structure and logical relationships explicit.

Other factors may include differences in model capability, explanation quality, and the diagnostic value of predicted probabilities. 
{Our findings reflect the particular LLM-generated explanations and occlusion-sensitivity heatmaps tested; other XAI methods, especially those tailored to users' needs, may yield different outcomes.}
RAVEN probabilities were often dispersed across answer options, whereas in LSAT the LLM frequently assigned very high predicted probabilities even when its selected answer was incorrect. {Generalizability is also limited by our university-student sample, whose responses may differ from those of domain experts in real-world settings. As AI models become more capable, the relative effectiveness of support strategies may change, although the importance of calibrated reliance and error recovery remains.} Because the studies differ simultaneously in task domain, model, explanation characteristics, and probability distributions, the present experiments cannot isolate the contribution of any single factor. {Future work should examine these factors across different models, explanation methods, and user populations.}

\subsection{Toward Adaptive Human--AI Support}

The RAVEN results also illustrate the potential value of selective automation. The derived policy used the gap between the two highest predicted probabilities to determine whether to accept the AI prediction or retain the participant's response, and achieved the highest descriptive accuracy. Because it reused responses from the predicted-probability condition rather than an independent participant group, it should be interpreted only as a reference benchmark. This approach is related to learning-to-defer frameworks~\cite{madras2018predict,mozannar2020consistent}.

Selective automation does not imply that AI must control when humans become involved. In many decision-support settings, humans retain final authority, and predicted probabilities may instead indicate when an AI recommendation warrants greater scrutiny. Uncertainty information also cannot reveal all failures: a model may be confidently incorrect, making explanations or other task-relevant evidence useful for identifying errors not reflected in its uncertainty.

More broadly, different forms of support provide different kinds of evidence. Their usefulness depends on the task, model, and reliability of the information they expose. Human--AI interfaces should therefore support calibrated reliance and effective error recovery by matching assistance to the decision context rather than adopting a single strategy universally.

\section{Conclusion}

Effective {AI-assisted decision-making} depends on matching
AI support to the task and to the evidence available for evaluating
model outputs. Across our studies, no single support strategy was
consistently best: predicted probabilities were descriptively
strongest in the visual reasoning task, while LLM-generated
explanations performed best in language-based logical reasoning.
{Rather than resolving the challenge of calibrated reliance,
this work offers an early foundation for understanding how different
forms of AI support shape human judgment across tasks. We hope
these findings serve as a useful reference point for future research
on designing AI-assisted decision-making systems that help users
recognize AI errors and exercise appropriate reliance.}

\begin{acks}
This research has been partially supported by the Israel Ministry of Innovation, Science \& Technology grant 1001818511 and the U.S. National Science Foundation grant CCF-1942836.
\end{acks}

\bibliographystyle{acm}

\clearpage
\appendix

\section{Participant and Human-Subject Procedures}
\label{app:participants}

\subsection{Recruitment and Demographics}
Participants in all three studies were recruited through university mailing lists and had academic backgrounds in STEM fields. All participants in the multi-stage RAVEN study were master's-degree students, whereas the between-subjects RAVEN and LSAT studies included both bachelor's- and master's-degree students. Sufficient English proficiency was a recruitment requirement, although proficiency was not formally assessed as part of the experiments. No participants were excluded from the two between-subjects studies; the multi-stage RAVEN study instead used the study-specific early-stopping procedure described in Appendix~B.1. Table~\ref{tab:demographics} summarizes the demographic characteristics of the analyzed samples.

\begin{table}[h]
\centering
\begin{tabular}{l|c|cc|cc}
\toprule
& & \multicolumn{2}{c|}{Gender} & \multicolumn{2}{c}{Age} \\
Study & Participants & Women & Men & Mean & SD \\
\midrule
RAVEN multi-stage        & 27  & 8  & 19 & 27.26 & 4.89 \\
RAVEN between-subjects   & 100 & 35 & 65 & 24.42 & 6.27 \\
LSAT between-subjects    & 80  & 26 & 54 & 28.25 & 6.73 \\
\bottomrule
\end{tabular}
\caption{Participant demographics across the three studies.}
\label{tab:demographics}
\end{table}

No a priori power analysis was conducted; accordingly, null findings, particularly in the smaller multi-stage RAVEN study, are interpreted cautiously.

\subsection{Inclusion and Exclusion Criteria}
Participants were included in the analyses only if they completed the
required experimental procedure for their assigned study and condition.
Participants who discontinued the study or did not complete the required
questions were excluded from the reported analyses.

\subsection{Informed Consent}
Before participating, all participants reviewed an informed consent form and voluntarily agreed to participate. They were informed that participation was voluntary and that they could discontinue at any time without penalty. Participants were also informed that demographic and experimental data would be treated confidentially, that demographic information including age and gender would be collected for research purposes, and that personally identifying information would not be disclosed in scientific publications.

\subsection{Performance-Based Compensation}
Across all three studies, compensation was performance-based: participants received 10 NIS for each correct answer, with a guaranteed minimum payment of 20 NIS. This payment scheme was identical across experimental conditions and provided a direct incentive for accurate responses.

\subsection{Task Item Selection}
Across the three studies, we intentionally selected relatively challenging
items so that participants would have a meaningful reason to consider the
AI-provided information rather than being able to solve most items immediately
without assistance.
For the first RAVEN study, 16 items were selected to provide broad coverage
across the different RAVEN configurations (see B.4) while favoring relatively
challenging problems.
For the second RAVEN study, the candidate pool included items from the first
study on which most participants had answered incorrectly, together with
additional challenging RAVEN items. The final items were selected so that the
displayed AI predictions were correct on six of the ten items and incorrect
on four, yielding an AI accuracy of 60\% across conditions. Model predictions
were not modified.
The LSAT study followed the same general rationale. Ten relatively challenging
questions were selected, and the model's responses to candidate items were
checked before data collection. The final set was chosen to contain six correct
and four incorrect model predictions, again maintaining an AI accuracy of
60\%.
This selection strategy was intended to avoid ceiling effects while ensuring
that the AI remained useful but fallible, creating meaningful opportunities
to examine how participants used the different forms of AI support.

\section{Additional Details of the RAVEN Multi-Stage Study}
\label{app:study1}

This appendix provides supplementary methodological and analysis
details for the RAVEN multi-stage study in
Section~\ref{sec:study1}. We first describe the study-specific
early-stopping procedure, followed by details on the AI models,
occlusion-sensitivity heatmaps, LLM explanation generation, and the
reported AI accuracy shown to participants. We then report
supplementary analyses of order effects and whether the effect of
explanations depended on the correctness of the preceding AI
prediction.

\subsection{Study Procedure and Early Stopping}
\label{app:study1-procedure}

Participants completed two explanation blocks, each containing up to
eight RAVEN puzzles. Within each block, we tracked disagreements between
the participant's initial response, provided before any AI information
was shown, and the model's predicted answer. The block terminated once
four such disagreements had occurred; otherwise, participants completed
all eight items. The disagreement counter was reset at the beginning of
the second block.

This stopping rule ensured that each block contained sufficient
opportunities to examine participants' responses when their initial
judgment differed from the AI recommendation, while limiting participant
fatigue. Thus, participants could complete at most 16 puzzles, but the
actual number varied depending on when the fourth disagreement occurred
within each block.

\subsection{RAVEN Models and Explanation Generation}

Two model sources were used in the study:
DCNet~\cite{zhuoeffective}, a CNN-based architecture trained on the
RAVEN dataset, and Claude 3.7 Sonnet. Within each explanation block,
the displayed prediction and its accompanying explanation came from
the same model source. In the LLM block, both the prediction and the
natural-language explanation were produced by Claude 3.7 Sonnet. In
the OS block, the prediction and occlusion-sensitivity visualization
were derived from DCNet. Thus, an explanation produced by one model
was never used to explain a prediction generated by the other model.

\subsection{Occlusion-Sensitivity Heatmap Generation}
\label{app:os-generation}

The occlusion-sensitivity heatmaps were generated using a
sliding-window occlusion procedure applied separately to each of the
16 panel images associated with a RAVEN item: eight context panels and
eight answer choices.

Each image was scanned using an occlusion window of size
\(20 \times 20\) pixels with a stride of 10 pixels, resulting in 50\%
overlap between adjacent windows. At each window position, the
corresponding image region was masked by setting its pixel values to
zero. We recorded the resulting absolute change in the model's output
score for its originally predicted answer,
\(\Delta = |s_{\mathrm{original}} - s_{\mathrm{occluded}}|\).

The importance values obtained at the different window positions were
accumulated across overlapping windows to form a pixel-level
sensitivity map. The resulting map was min--max normalized to the
interval \([0,1]\). For presentation to participants, the normalized
heatmap was rendered using a jet colormap and alpha-blended with the
original grayscale image using equal image and heatmap weights.

\subsection{RAVEN Problem Configurations}

The RAVEN dataset organizes Raven's Progressive Matrix problems into
seven predefined figural configurations that differ in the spatial
organization and nesting of visual elements \cite{zhang2019raven}.
These configurations determine how objects are arranged within each
matrix panel, while the underlying relational rules may vary across
individual problems within the same configuration.
The seven configurations are:
\begin{itemize}
    \item \texttt{center\_single}: a single visual element positioned
    at the center of the panel.
    \item \texttt{distribute\_four}: multiple elements arranged in a
    four-position spatial layout.
    \item \texttt{distribute\_nine}: multiple elements arranged in a
    nine-position spatial layout.
    \item \texttt{left\_center\_single\_right\_center\_single}:
    two spatial components positioned on the left and right, each
    containing a centered element.
    \item \texttt{up\_center\_single\_down\_center\_single}:
    two spatial components positioned above and below one another,
    each containing a centered element.
    \item \texttt{in\_center\_single\_out\_center\_single}:
    a nested configuration containing an inner centered element and
    an outer centered element.
    \item \texttt{in\_distribute\_four\_out\_center\_single}:
    a nested configuration in which the inner component contains
    multiple spatially distributed elements and the outer component
    contains a centered element.
\end{itemize}

\subsection{RAVEN LLM Prompt and Generation Procedure}

\begin{tcolorbox}[
    colback=gray!5,
    colframe=gray!40,
    title={RAVEN base prompt},
    fonttitle=\bfseries,
    breakable
]
Here is a question from the RAVEN-10k dataset and the choices.
Please provide an explanation to the best choice to your opinion
and provide 3--4 rows of the logic behind.
\end{tcolorbox}

The same base instruction was used across most RAVEN items, with only the puzzle image and answer choices varying between items.
For some items, the model showed difficulty in reliably counting visual elements. In these cases, the prompt was augmented with a brief numerical description of the relevant element counts to help the model attend to the quantitative structure of the puzzle rather than relying on visual counting alone.

The explanations were generated through the Claude conversational
interface rather than through an API. Sampling parameters such as
temperature and top-\(p\) were not exposed by the interface and were
therefore neither manually specified nor recorded. The explanations
were presented to participants as generated, without editing,
shortening, or other post-processing.

\subsection{Reported AI Accuracy}
\label{app:reported-accuracy}

Participants were informed that the AI system had an approximate
historical accuracy of 76\%. This value was used as a coarse aggregate
summary of two independent model evaluations. DCNet achieved an
average accuracy of 87.9\% on the RAVEN test data, whereas Claude
3.7 Sonnet correctly solved 10 of 16 RAVEN items, corresponding to
62.5\%. The aggregate figure was intended to communicate that the AI
system was capable but fallible, rather than to provide a calibrated
per-item probability.

\subsection{Exploratory Order Effect}
\label{app:order-effect}

We observed an exploratory order-dependent pattern in the multi-stage
study. Participants who began with LLM-generated explanations achieved
higher accuracy early in the task than those who began with OS
explanations (43.1\% vs.\ 20.0\%). Given the small sample size and the
exploratory nature of this comparison, we do not draw strong
conclusions from this result.

\subsection{Moderation by AI-Prediction Correctness}
\label{app:moderation}

Because AI-prediction correctness was determined before explanation
exposure, we examined whether it moderated the incremental effect of
the explanation.

We fitted a logistic generalized estimating equation model to 206
complete paired trials from 27 participants. The model predicted
participant correctness from reveal stage, AI-prediction correctness,
explanation type, their interactions, and pre-AI correctness, with
repeated observations clustered by participant using an exchangeable
working-correlation structure and robust standard errors.

Aggregating across explanation types, participant accuracy was
unchanged after explanation exposure when the AI prediction was
incorrect (33.67\% after both the prediction and the explanation) and
decreased slightly when the AI prediction was correct (64.81\% to
62.04\%). The Stage \(\times\) AI-correctness interaction was not
significant (\(OR=0.83\), 95\% CI \([0.37,1.86]\), \(p=.648\)),
providing no evidence that the effect of explanations depended on
whether the preceding AI prediction was correct. The Stage
\(\times\) AI-correctness \(\times\) explanation-type interaction was
also not significant (\(OR=0.79\), 95\% CI \([0.22,2.90]\),
\(p=.723\)).

Thus, the results do not provide evidence that explanations become
more beneficial specifically when AI predictions are incorrect. Given
the relatively small sample in this study, these null interaction
results should be interpreted cautiously.

\subsection{Exploratory Analysis of Initial Human Confidence}
\label{app:confidence}

{We conducted a post hoc exploratory analysis of whether participants' confidence in their initial answers was associated with their reliance on AI recommendations in the multi-stage RAVEN study. The analysis included 206 complete paired trials from 27 participants in which the initial human answer differed from the AI prediction. For these trials, adopting the AI-predicted answer provides a direct measure of reliance on the AI recommendation.}

{We first examined whether pre-AI confidence (rated on a five-point scale) predicted adoption of the AI answer after the prediction stage. We fitted a logistic generalized estimating equation (GEE) model clustered by participant, controlling for AI-prediction correctness, initial human-answer correctness, and support condition (LLM versus OS). Higher initial confidence was significantly associated with lower odds of adopting the AI recommendation (OR = 0.58, 95\% CI [0.45, 0.75], $p < .001$). Descriptively, the adoption rate decreased from 81.8\% at confidence level 1 to 35.9\% at confidence level 5.}

{We then examined whether the association between initial confidence and reliance differed across support conditions. Using adoption of the AI answer after the explanation stage as the outcome, we fitted a participant-clustered logistic GEE model with a confidence-by-support-condition interaction, controlling for AI-prediction correctness and initial human-answer correctness. The interaction was not statistically significant (OR = 0.94, 95\% CI [0.56, 1.59], $p = .821$), providing no evidence that the association between initial confidence and reliance differed between the LLM and OS conditions.}

\section{Additional Details of the RAVEN Between-Subjects Study}
\label{app:study2}

This section provides additional methodological details for the RAVEN between-subjects study, including the sources of the AI predictions across conditions, the generation of predicted probabilities, and the derived selective-automation benchmark.

\subsection{AI Prediction Sources and Item Selection Across Conditions}

The source of the AI prediction varied across the four AI-support conditions. To hold AI accuracy constant across conditions, we first obtained model predictions for candidate RAVEN items and then selected 10 items for each condition such that six displayed predictions were correct and four were incorrect. Model predictions were not modified to achieve this distribution; correctness was controlled through the selection of task items.

In the LLM-explanation condition, Claude 3.7 Sonnet was used for all 10 selected RAVEN items, with six correct and four incorrect predictions; the accompanying natural-language explanations were generated by the same model. In the OS-heatmap condition, DCNet was used for all 10 selected items, likewise yielding six correct and four incorrect predictions, with the occlusion-sensitivity heatmaps derived from the same model.

The Prediction Only condition combined predictions from both model sources. Five selected items displayed predictions from Claude 3.7 Sonnet, consisting of three correct and two incorrect predictions, while the other five displayed predictions from DCNet, also consisting of three correct and two incorrect predictions. This produced the same overall composition of six correct and four incorrect AI predictions while balancing the two model sources.

In the predicted-probability condition, both the predicted answer and the predicted probability distribution over the eight answer choices were derived from DCNet. The 10 selected items again consisted of six correct and four incorrect DCNet predictions.

\subsection{Generation of Predicted Probabilities}

For each RAVEN item, DCNet received the 16 panel images as a single input and produced eight raw output logits, one for each answer choice. These logits were converted using the softmax function into eight normalized values that summed to one and were displayed to participants as the predicted probability distribution across the answer choices. The answer choice with the largest logit was used as the AI's predicted answer.

No additional calibration procedure, such as temperature scaling, was applied. The displayed predicted probabilities should therefore be interpreted as the softmax-normalized outputs of DCNet rather than as calibrated estimates of the model's empirical likelihood of being correct.

\subsection{Derived Selective-Automation Policy}
\label{app:selective-policy}

In addition to the five participant-facing conditions, we evaluated a selective-automation policy as a derived theoretical benchmark. This policy did not involve an additional participant group.

For each item, the policy compared the two largest predicted probabilities produced by DCNet. If the probability gap between the highest- and second-highest-ranked answer choices exceeded 10 percentage points, the policy selected the AI's predicted answer. Otherwise, it retained the participant's answer.

The 10-percentage-point threshold was specified before participant performance was examined. The policy was evaluated using the responses of the same 20 participants assigned to the predicted-probability condition. It therefore does not constitute an independent experimental condition, and its results are reported as a descriptive reference benchmark rather than as an independent group in between-subjects statistical comparisons.

\section{Additional Details of the LSAT Between-Subjects Study}
\label{app:study3}

\subsection{Representative User Interfaces}

\begin{figure}
\centering
\begin{subfigure}[b]{0.48\textwidth}
\centering
\includegraphics[width=\textwidth]{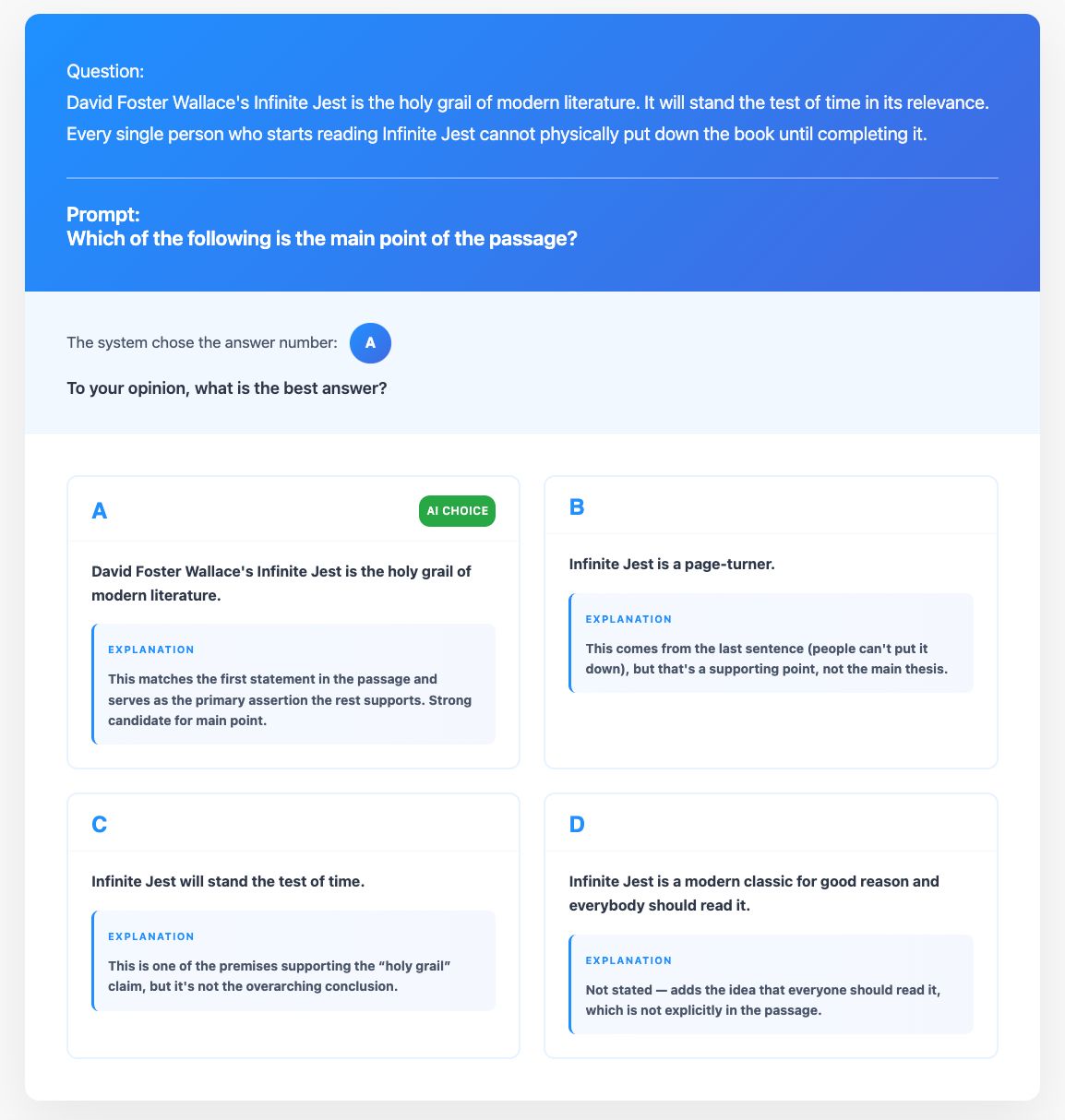}
\caption{Prediction + LLM Explanation}
\end{subfigure}
\hfill
\begin{subfigure}[b]{0.48\textwidth}
\centering
\includegraphics[width=\textwidth]{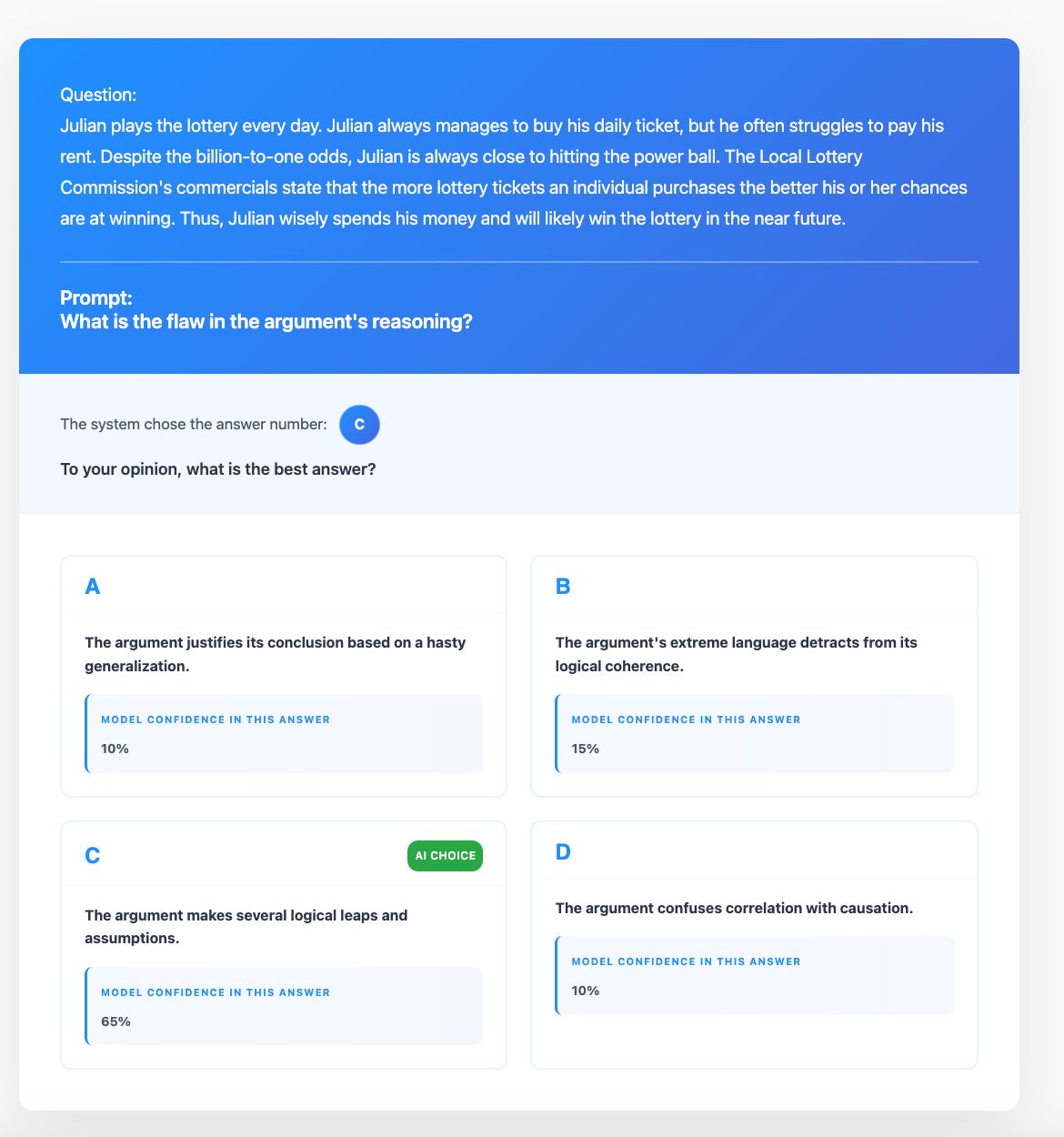}
\caption{Prediction + Probability}
\end{subfigure}
\caption{Illustrative user interface examples from the LSAT study.}
\label{fig:lsat_interface}
\end{figure}

Figure~\ref{fig:lsat_interface} shows representative interfaces from the Prediction + LLM Explanation and Prediction + Probability conditions. These examples illustrate how the AI prediction and accompanying support information were presented to participants during the LSAT study.

\subsection{LLM Predictions, Explanations, and Predicted Probabilities}

The LSAT predictions, explanations, and predicted probabilities were generated using ChatGPT-5 through OpenAI's conversational interface rather than through an API.

\begin{tcolorbox}[
    colback=gray!5,
    colframe=gray!40,
    title={LSAT explanation and probability prompt},
    fonttitle=\bfseries,
    breakable
]
You are an expert in LSAT questions. Here is an LSAT question and four answer choices (A, B, C, D). For each answer choice, give me a short explanation (1--2 sentences) describing what it says and how it relates to the question or argument and your confidence for each choice to be the correct answer.
\end{tcolorbox}

For consistency with the terminology used throughout the paper, we refer to the numerical values returned for the answer options as \emph{predicted probabilities}. These values were elicited directly from the model through the prompt above rather than computed from token-level logits. Because the conversational interface did not expose token-level log-probabilities, we could not independently derive the probabilities from the model's internal output scores. No additional calibration procedure was applied.

Sampling parameters such as temperature and top-\(p\) were not exposed by the conversational interface and were therefore neither manually specified nor recorded. The generated explanations and predicted probabilities were presented to participants without editing or other post-processing.

\subsection{Source of Expert Explanations}

The LSAT expert explanations were adopted from Bansal et al.~\cite{bansal2021does} and were based on a commercially published LSAT preparation guide. In the original study, explanations for the correct answers were condensed to a maximum of two sentences by one of the authors. Because the preparation guide did not provide explanations for the incorrect answer choices, the original authors manually created explanations for the alternative choices while matching the tone and conciseness of the correct-answer explanations.
These expert explanations were used only in the \emph{Prediction + Expert Explanation} condition. They did not determine the AI predictions or the predicted probabilities presented in the other conditions.

\end{document}